\documentclass[prl,twocolumn,amsmath,amssymb,superscriptaddress,nofootinbib,aps,10pt]{revtex4-1}

\usepackage{amssymb}
\usepackage{bbold}
\usepackage{xcolor}
\usepackage{makecell}
\usepackage{enumerate}
\usepackage[shortlabels]{enumitem}
\usepackage{graphicx}
\usepackage[caption=false]{subfig}
\usepackage[T1]{fontenc}

\usepackage{hyperref}
\usepackage[capitalise]{cleveref}

\begin{document}

\title{Cell proliferation maintains cell area polydispersity in the growing fruit fly wing epithelium}

\author{Michael F. Staddon}
\affiliation{Max Planck Institute for Physics of Complex Systems, Dresden, Germany}
\affiliation{Center for Systems Biology, Dresden, Germany}
\affiliation{Max Planck Institute of Molecular Cell Biology and Genetics, Dresden, Germany}
\affiliation{Cluster of Excellence, Physics of Life, TU-Dresden, Dresden, Germany}
\author{Natalie A. Dye}
\affiliation{Cluster of Excellence, Physics of Life, TU-Dresden, Dresden, Germany}
\affiliation{Mechanobiology Institute, National University of Singapore, Singapore}
\affiliation{Biomedical Engineering Department, National University of Singapore, Singapore}
\author{Marko Popovi\'c}
\email{mpopovic@pks.mpg.de}
\email{popovic@irb.hr}
\affiliation{Max Planck Institute for Physics of Complex Systems, Dresden, Germany}
\affiliation{Center for Systems Biology, Dresden, Germany}
\affiliation{Cluster of Excellence, Physics of Life, TU-Dresden, Dresden, Germany}
\affiliation{Ru\dj er Bo\v skovi\'c Institute, Zagreb, Croatia}
\author{Frank Jülicher}
\email{juelicher@pks.mpg.de}
\affiliation{Max Planck Institute for Physics of Complex Systems, Dresden, Germany}
\affiliation{Center for Systems Biology, Dresden, Germany}
\affiliation{Cluster of Excellence, Physics of Life, TU-Dresden, Dresden, Germany}

\begin{abstract}
Developing epithelial tissues coordinate cell proliferation and mechanical forces to achieve proper size and shape. As epithelial cells tightly adhere together to form the confluent tissue, the distribution of cell areas significantly influences possible patterns of cellular packing and thereby also the mechanics of the epithelium. Therefore, it is important to understand the origin of cell area heterogeneity in developing tissues and, if possible, how to control it. Previous models of cell growth and division have been successful in accounting for experimentally observed area distributions in cultured cells and bacterial colonies, but developing tissues present additional complexity due to self-organized patterns of mechanical stresses that guide morphogenesis. Here, we address this challenge focusing on the \textit{D. melanogaster} wing disc epithelium. We consider a simple model that couples cell cycle dynamics to tissue mechanics. From time-lapse imaging of the cellular network, we extract all model parameters — cell growth rates, division rates, and mechanical fluctuations — revealing that they all depend on cell size. With these independently measured parameters, our model quantitatively reproduces the observed cell area distribution without any fitting parameters and further predicts tissue pressure gradients, in quantitative agreement with previously published data. Importantly, we find that cell proliferation accounts for 85\% of cell area variance, establishing it as the dominant source of packing disorder that influences tissue mechanics and organization.
\end{abstract}

\maketitle

\section{Introduction}
Many biological processes, such as morphogenesis \cite{friedl2009collective, behrndt2012forces, etournay2015interplay, julicher2017emergence, maniou2021hindbrain}, cancer progression \cite{friedl2012classifying, arwert2012epithelial}, and wound healing \cite{brugues2014forces, tetley2019tissue, ajeti2019wound}, involve the coordinated dynamics of cells that drives tissue remodelling. This behaviour is regulated by both active stresses generated within the tissue \cite{salbreux2012actin, murrell2015forcing, ladoux2017mechanobiology} and the mechanical properties of the tissues, which govern how effectively cells can transmit forces or rearrange and change their neighbors. 
Cell growth and division can act as a source of active stress \cite{firmino2016cell,  devany2021cell,Tahaei2024}, generating a mechanical stress field in the tissue, which in turn can regulate tissue mechanical properties, for example by triggering cell rearrangements and fluidising the tissue \cite{ranft2010fluidization, godard2019cell, bocanegra2022cell}. 
Moreover, tissue mechanics can influence the rate and orientation of cell divisions, which in turn can affect the tissue mechanics. 
For example, cells typically divide along their long axis, which is a result of overall tissue mechanics, generating two more isotropic daughter cells, thereby reducing tissue stress \cite{wyatt2015emergence}.


The cycle of cell growth and division also regulates the distribution of cell sizes, or the size polydispersity, together with tissue-scale pressure and short time-scale fluctuations in active stresses. Cell sizes within bacterial populations \cite{hirano1982lognormal, loper1984lognormal, koutsoumanis2013stochasticity}, and also within epithelial tissues \cite{zehnder2015cell, puliafito2017cell}, have been reported to follow log-normal distributions. In the context of epithelial tissues, cell size is represented by the projected cell area on the tissue surface, which we also employ here. The distributions appear self-similar: cell size distributions for different cell types follow the same distribution function up to a scale factor. These distributions can be described using models incorporating exponential growth and normally distributed division times or growth rates, either between cells or across the cell's lifetime \cite{koutsoumanis2013stochasticity, amir2014cell}, or alternatively with cells dividing stochastically at some rate \cite{hosoda2011origin, wilk2014universal, puliafito2017cell, genthon2022analytical}.
While such models have successfully explained cell area distributions in cultured cells \cite{wilk2014universal}, developing tissues present additional complexity: mechanical stresses vary spatially as tissues prepare for or undergo morphogenetic events, and it remains unclear how cell cycle dynamics interact with tissue mechanics to shape cell area distributions. An example of such mechanical heterogeneity is the wing disc pouch of the fruit fly, which exhibits a radial gradient of in-plane pressure and a corresponding gradient of cell areas \cite{dye2021self}. Understanding cell area distributions in this context requires accounting for both cell cycle dynamics and tissue mechanics.

In this work, we study cell area heterogeneity in the wing disc epithelium of the developing fruit fly \textit{D. melanogaster}. We directly measure cell growth rates, division rates, and mechanical fluctuations from time-lapse imaging, discovering that these quantities depend on cell size. We develop a model that integrates these cell cycle dynamics with tissue mechanics through pressure fluctuations, reproducing the observed cell area distribution without any fitting parameters. 
Our results show that cell proliferation contributes about $85\%$ to the overall cell area variance, with mechanical fluctuations accounting for the remainder. Furthermore, by incorporating spatially varying pressure, our model predicts tissue pressure gradients that agree quantitatively with independent laser ablation measurements \cite{dye2021self}. This demonstrates that cell area distributions in mechanically complex tissues encode information about both cell cycle dynamics and tissue mechanics.
The analysis presented in this work can be extended to other biological tissues and provides a framework for understanding how proliferation and mechanics together shape tissue organization.

\section*{Statistics and dynamics of cell areas in the developing fly wing pouch}

\begin{figure}
    \centering
    \includegraphics[width=.5\textwidth]{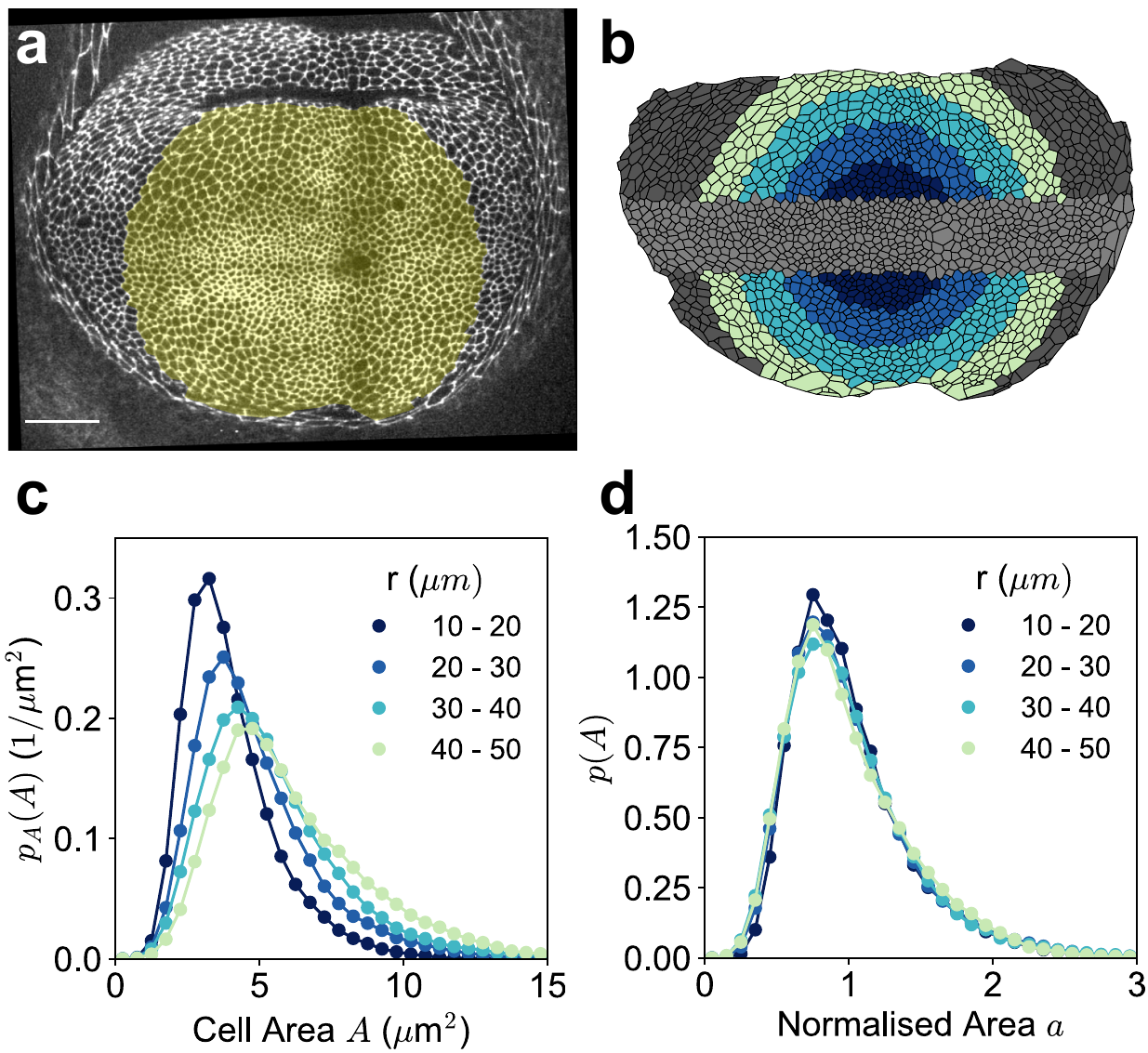}
    \caption{Cell area distribution in the \textit{Drosophila} wing imaginal disc. (a) Image of the \textit{Drosophila} wing disc epithelium. The yellow circle highlights the region of the wing disc analysed. The scale bar shows 100 $\mu m$. (b) Segmented cells are divided in radial bins based on their distance from the center of the tissue \cite{dye2021self}, as indicated by different colors. Light gray cells are not analysed as they belong to the DV boundary region and have been shown to have different material properties \cite{dye2021self}. Dark gray cells are outside the region of interest. (c) Probability density of cell areas by distance from the center, for bins of width 2 $\mu m^2$. (d) Probability density of cell area normalised by mean area within the bin, for bins of width 0.4.}
    \label{fig:1}
\end{figure}

We study the heterogeneity of cell areas in the developing fruit fly wing imaginal disc in the late larval stage, which is the precursor of the adult wing tissue. 
We analyse previously published data from time-lapse experiments where explanted wing imaginal discs were imaged for about $13$ hours \cite{dye2021self}. 
Since we are interested in the intrinsic heterogeneity of cell areas, we aim to correct for systematic contributions to the cell area heterogeneity that stem from spatial gradients of mechanical stresses and cell mechanical properties. 
In Ref. \cite{dye2021self} a radial gradient of cell areas around the center of the cell pouch was reported, and cells in the vicinity of the dorso-ventral (DV) compartment boundary were inferred from laser ablation experiments to have significantly different elastic properties compared to cells in other tissue regions. 
Therefore, for our analysis we divide the pouch region in radial bins around the center position identified in Ref. \cite{dye2021self} (colored annuli in Fig.\ref{fig:1}~(b)) excluding a stripe of cells near the DV compartment boundary (light gray stripe Fig.\ref{fig:1}~(b)).

We measure the cell area distribution in each radial bin and find that the mean cell area increases with the distance from the center, see Fig.~\ref{fig:1}(c), consistent with  \cite{dye2021self}.  
Strikingly, when we normalise cell area $A$ by the average cell area in the corresponding radial bin $\overline{A}(r)$, we find that the cell area distributions collapse on a universal distribution function. Here, the average $\overline{A}(r)$ is calculated from all cells in the bin $r$ throughout the experiment. This suggests that the normalised cell area $a = A/\overline{A}(r)$ is the relevant variable to consider.
Can we understand the origin of this universality in the wing disc pouch? 

\begin{figure}
    \centering
    \includegraphics[width=.5\textwidth]{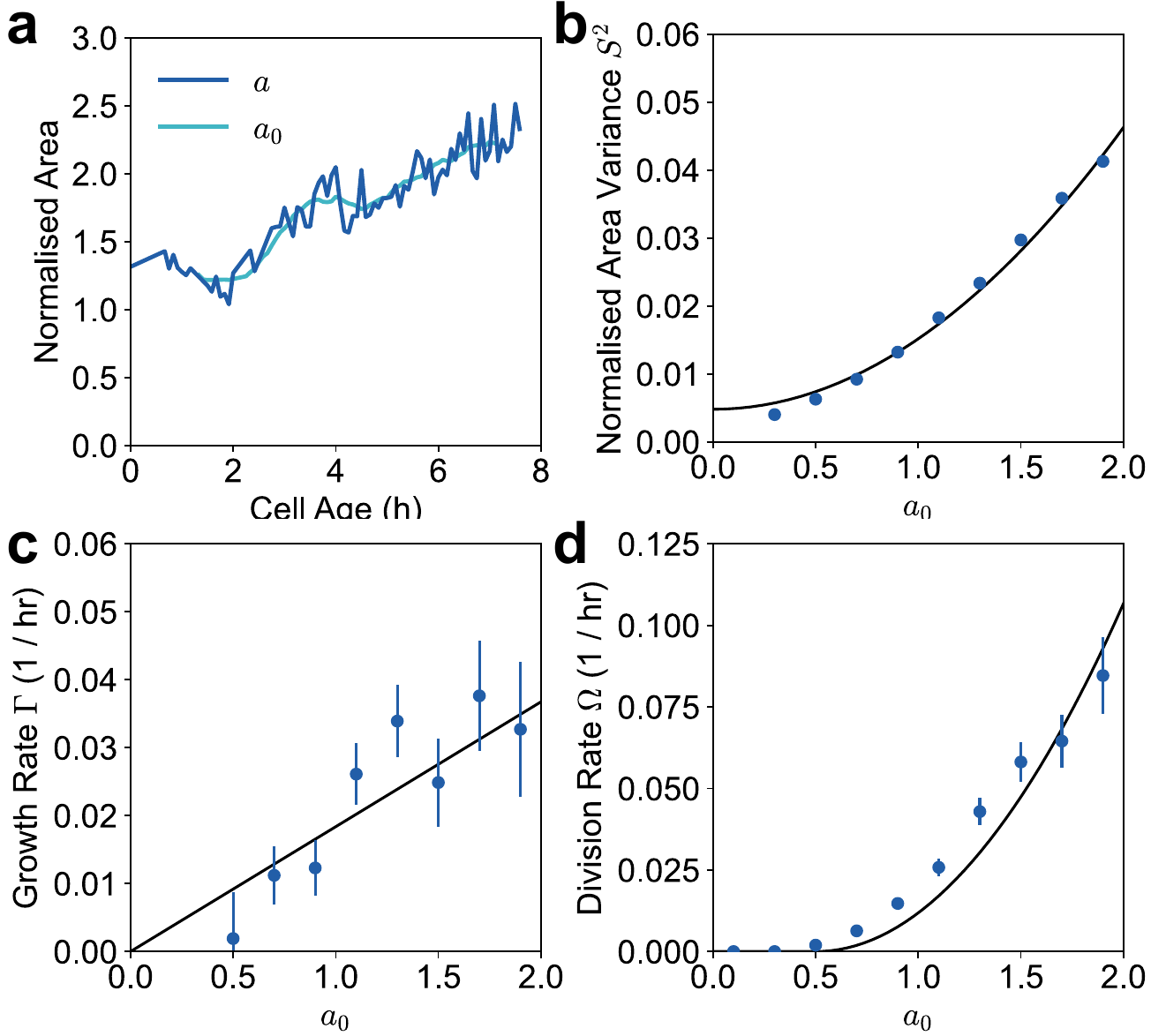}
    \caption{Quantification of cell area dynamics: growth, division and noise. (a) An example cell area trajectory over time. Dark blue line shows the normalised cell area $a(t)$. Light blue line is the one hour moving average of the cell area $a_{c,0}(t)$. (b) Normalised cell area variance in a $1\text{hr}$ moving time-window $S^2$ against the cell target area $a_0$, defined as the average cell area in the moving window. Data points show the mean values from experiments $\pm$ standard error binned by $a_0$ in bins of width 0.2. Solid line is the quadratic function $s^2(a_0)$ we use to represent the data. (c) Cell growth rate $\Gamma$ against $a_0$. Data points show the mean values from experiments $\pm$ standard error binned by $a_0$. Solid line is the linear function $\gamma(a_0)$ used to represent the data in the model. (d) Cell division rate $\Omega$ against $a_0$. Data points show the mean values from experiments $\pm$ standard error binned by $a_0$. Solid line shows the function $\omega(a_0)$. For each statistic, n=296036 data points were used, consisting of 3686 unique cells across 4 experiments.}
    \label{fig:2}
\end{figure}

We start by analysing the dynamics of the normalised cell area in individual cells. A typical trajectory of a single normalised cell area over time $a(t)$ is shown in Fig.~\ref{fig:2}(a). We observe that normalised cell area steadily grows over multiple hours. However, we note that cell area also fluctuates on the time-scale of minutes, likely as a consequence of mechanical noise generated in the cell and its surroundings. Furthermore, when a cell divides it produces two daughter cells, each approximately half the area of the mother cell. All these processes contribute to the heterogeneity of the cell areas in the tissue. Our goal is to determine whether the observed growth, divisions and noise are sufficient to account for the observed universal distribution of cell areas.

\begin{figure*}
    \centering
    \includegraphics[width=1\textwidth]{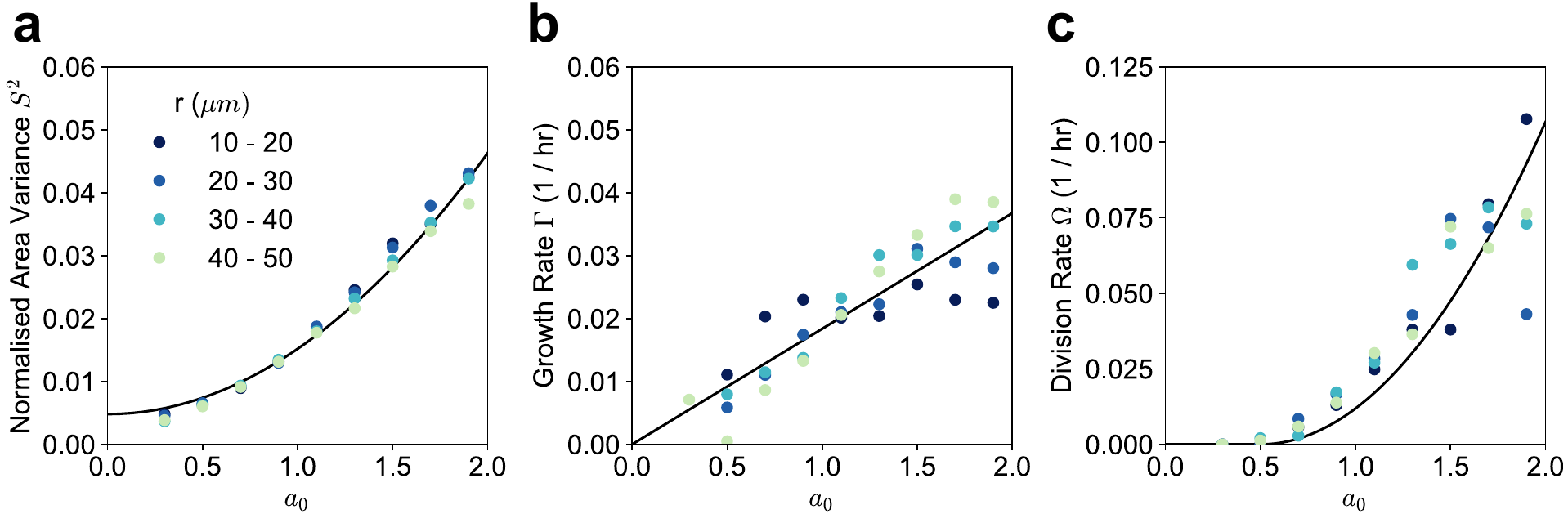}
    \caption{(a) Normalised cell area variance in a $1\text{hr}$ moving time-window $S^2$ against the cell target area $a_0$, defined as the average cell area in the moving window. Data points show the mean values from experiments binned by $a_0$ in bins of width 0.2. Different colors correspond to radial distances from the wing disc center. Solid line is the quadratic function $s^2(a_0)$ we use to represent the data in the main text. (b) Cell growth rate $\Gamma$ against $a_0$. Data points show the mean values from experiments binned by $a_0$. Different colors correspond to radial distances from the wing disc center. Solid line is the linear function $\gamma(a_0)$ used to represent the data in the main text. (c) Cell division rate $\Omega$ against $a_0$. Data points show the mean values from experiments binned by $a_0$. Different colors correspond to radial distances from the wing disc center. Solid line shows the function $\omega(a_0)$ we use to represent the data in the main text.}
    \label{sfig:1}
\end{figure*}

We now quantitatively describe these three processes from the experimental data. To quantify the normalised cell area fluctuations, we determine for each cell the average $a_{0}$ in a moving window of duration $1\text{hr}$, which we denote as the cell target area. We chose the $1\text{hr}$ time-window as a scale much shorter than the cell cycle and still longer than minutes time-scale of cell fluctuations. The corresponding variance is denoted $S^2$. We show an example of cell target area $a_{0}$ in Fig.~\ref{fig:2}(a) together with the normalised cell area $a$. We then quantify the average variance $\bar{S}^2(a_0)$, see Fig.~\ref{fig:2}(b), where the average is taken over all cells and over time, binned by cell target area $a_0$. We find that the magnitude of cell area fluctuation grows with the cell target area, which can be represented by a quadratic function
$s^2(a_0) = c_s a_0^2 + d_s$
where we obtain $c_s=(1.037 \pm 0.002)\cdot 10^{-2}$ and $d_s=(4.80 \pm 0.04)\cdot 10^{-3}$.
We verified that this relationship holds in individual radial bins as well, see Supp. Fig.\ref{sfig:1}(a) . 

We measure the cell growth rate as the rate of relative change of the cell target area. Therefore, for a cell at time $t$, the growth rate is defined as $\Gamma(t) = (a_{0}(t + \Delta t) - a_{0}(t))/(a_{0}(t) \Delta t)$ where $\Delta t$ is the interval between two subsequent time-frames. 
We show the average cell growth rate $\bar{\Gamma}(a_0)$ as a function of the target area in Fig.~\ref{fig:2}(c). We can represent the average cell growth rate by a linear function $\gamma(a_0) = \alpha a_0$, where $\alpha = (0.0184 \pm 0.0003) \text{hr}^{-1}$. This expression applies to cells in individual radial bins as well, see Supp. Fig.\ref{sfig:1}(b).
It is interesting to note that this corresponds to a super-exponential growth. 

Finally, we quantified the cell division rate $\Omega$ by counting the number of cell divisions per unit time, binned by the cell target area $a_0$.  
We find almost no divisions of cells with normalised cell area less than $0.5$, and the division rate of larger cells increases approximately quadratically with cell target area (Fig.~\ref{fig:2}(d)). Therefore, we represent the measured cell division rate by a function
$\omega(a_0) = \theta(a_0 - 0.5) c_d (a_0 - 0.5)^2$, where $\theta$ is the Heaviside theta function and $c_d= (0.0420 \pm 0.0013 )\text{hr}^{-1}$.
This trend remains consistent when considering cells in individual radial bins, see Supp. Fig.\ref{sfig:1}(c).


\section*{Theory of cell area in a growing epithelium}

Having developed a quantitative description of the statistical properties of individual cell behaviours, we now build a simple stochastic model of cells in a proliferating tissue that incorporates the experimentally measured growth rate, division rate, and fluctuations.

\begin{figure}
    \centering
    \includegraphics[width=0.5\textwidth]{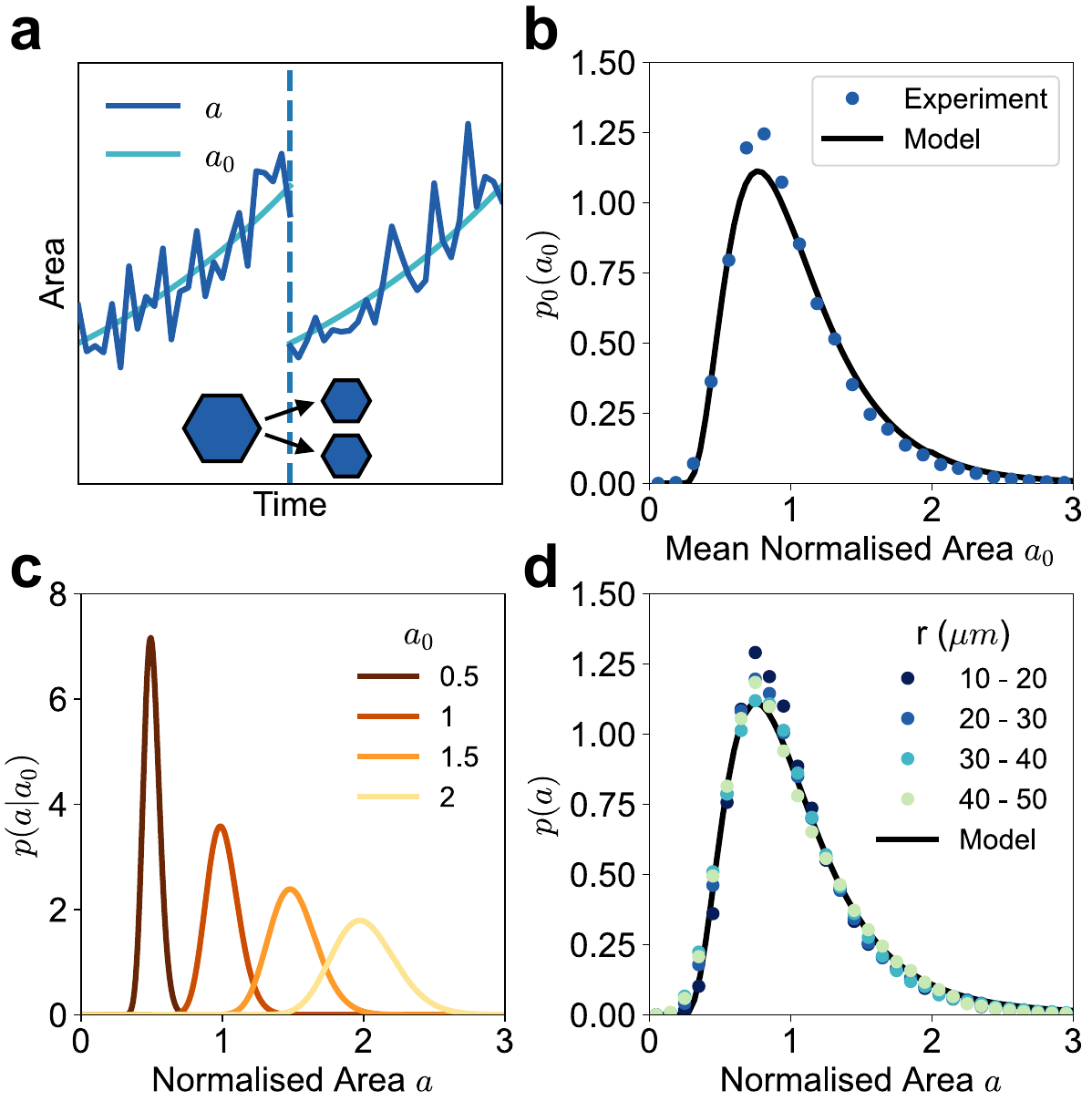}
    \caption{Cell area theory captures the experimental distribution. (a) Schematic showing a model cell area trajectory. The target area grows and can halve at the time of cell division for each of the daughter cells, while the actual area fluctuates about the target area. (b) Normalised target area probability density $p_0(a_0)$ against target area $a_0$ for different growth rates. (c) Conditional probability density of normalised cell area $p(a|a_0)$ given different target areas $a_0$. (d) Normalised cell area distribution $p(a)$. Dots show experimental data. The line shows the model prediction.}
    \label{fig:3}
\end{figure}

Motivated by the experimental observation that cell areas grow on a time-scale of hours and exhibit fluctuations on a time-scale of minutes, we  introduce a model that captures target cell area growth and pressure induced cell area fluctuations. We express the decomposition of the normalized cell area as $a = a_0 + \delta a$, where $a_0$ is the cell target area in absence of mechanical fluctuations, which increases over time due to cell growth, and cell area fluctuation $\delta a$ around the target value, as schematically illustrated in Fig.~\ref{fig:3}(a).  

We first focus on the target cell area $a_0$ for which we choose the growth rate to be the function $\gamma(a_0)$ we use to represent the experimentally measured average growth rate in Fig.~\ref{fig:2}(c): $\partial_t a_0= \gamma(a_0) a_0$. 
We describe the cell division as a stochastic process with division rate $\omega(a_0)$, which we take to be the function that represents the experimentally measured division rate in Fig.~\ref{fig:2}(d). The growth and division rules fully define the dynamics of $a_0$ and we determine its steady state distribution $p_0(a_0)$ from the equation
\begin{align}
    \label{eq:p0}
    \begin{split}
    \partial_{a_0}(\gamma(a_0) a_0 p_0(a_0)) &= 2 \omega(2 a_0) p_0(2 a_0) \\ &- \omega(a_0)p_0(a_0) - \lambda p_0(a_0)
    \end{split}
\end{align}
where the left hand side represents cell growth, and the right hand side represents cell divisions during which a cell of target area $2 a_0$ divides into the two daughter cells of target area $a_0$. Finally, $\lambda$ is a factor ensuring normalisation of the probability $\int p_0(a_0)d a_0 = 1$. We consider the steady state as relevant because our data is taken from the late stage of larval wing development, during which number of cells increases by a factor of about $1000$ \cite{Buchmann2014}.  We solve Eq. \ref{eq:p0} numerically, with boundary conditions $p_0(0) = p_0(+\infty) = 0$, as described in Methods, and we show the result in Fig.~\ref{fig:3}(b) together with the experimentally measured distribution $P(a_0)$. 
Notably, we find that the variance of the target area distribution, $\mathbb{V}ar[a_0] = 0.216$, is 85\% of the experimentally measured normalised area distribution, $\mathbb{V}ar[a] = 0.252$, suggesting that cell cycle dynamics are responsible for the majority of the observed area distribution.

To fully account for the observed cell area statistics we introduce two-dimensional pressure fluctuation $\Pi$ 
experienced by individual cells (Fig.~\ref{fig:3}(a)) that give rise to the cell area fluctuations $\delta a$ around $a_0$. Pressure fluctuations are induced by stochastic kicks $\xi$ originating from active mechanics of the cell and its environment, and fluctuations relax on a time-scale $\tau$  
\begin{equation}
    \partial_t \Pi =  - \frac{1}{\tau} \Pi  + \xi \quad.
\end{equation}
For simplicity, we consider $\xi$ to be a Gaussian random variable with $\mathbb{E}[\xi(t)] = 0$ and $\mathbb{E}[\xi(t)\xi(t')] = D \delta(t - t')$, where $D$ sets the magnitude of the noise. 
Therefore, the probability density of the pressure, $p_\Pi(\Pi, t)$ obeys the equation
\begin{equation}
     \partial_t p_\Pi(\Pi, t) = D \partial^2_\Pi p_\Pi + \frac{1}{\tau}\partial_\Pi (\Pi p_\Pi)
\end{equation}
where the first term on the right hand side represents the diffusion of probability due to mechanical noise, and the second term represents the relaxation. Steady state solution of this equation is a normal distribution
\begin{equation}
    p_\Pi(\Pi) = \frac{1}{\sqrt{2 \pi D \tau}} e^{-\frac{1}{2 D \tau}\Pi^2}
\end{equation}
with mean $0$ and variance $D \tau$. 

To relate the fluctuations of cell area and pressure, we propose a constitutive relation $\Pi= -K \ln{(a/a_0)}$, where the pressure fluctuation is proportional to the cell area logarithmic strain. 
Together with the distribution of pressure fluctuations, this relation allows us to calculate the conditional distribution
\begin{equation}
    \label{eq:p|a0}
    p(a | a_0) = \frac{K}{\sqrt{2 \pi D \tau}}\frac{a_0}{a} e^{-\frac{1}{2D\tau} \left(K \ln{\left(\frac{a}{a_0}\right)}\right)^2} .
\end{equation}
This is a log-normal distribution with parameters $\mu = \ln{a_0}$ and $\sigma= \sqrt{D \tau}/K$, which specify the mean $\mathbb{E}(a|a_0)= \exp{(\mu + \sigma^2/2)}$ and the variance $\mathbb{V}ar[a|a_0]= (\exp{(\sigma^2)} - 1) \mathbb{E}[a|a_0]^2$.  The conditional variance is the analogue of the experimentally measured normalised cell area variance as a function of $a_0$, which we represented in the previous section by the function $s^2(a_0)$ and showed in Fig.~\ref{fig:2}(b). Therefore, using the expressions for the mean and the variance of the log-normal distribution in Eq. \ref{eq:p|a0}, we determine values of $\sigma(a_0)$ at each value of $a_0$ for which $s^2(a_0)$ was evaluated. 

We now determine the normalised cell area distribution as the integral 
\begin{equation}
    \label{eq:p}
    p(a) = \int_0^\infty p(a | a_0) p_0(a_0) d a_0.
\end{equation}
It is important to notice that all model parameters are measured directly and no free parameters are needed.

We obtain an excellent agreement between the normalised cell area distribution predicted by our model and the actual experimental distribution of normalised cell areas, with an $R^2$-accuracy of $97.1 \pm 0.2 \%$ (Fig.~\ref{fig:3}(d)). Therefore, the universal distribution of the normalised cell area in the fruit fly wing disc can be understood as the result of cell proliferation, which includes cell growth and division, and mechanical noise manifested in pressure fluctuations.

\begin{figure*}[ht!]
    \centering
    \includegraphics[width=1\textwidth]{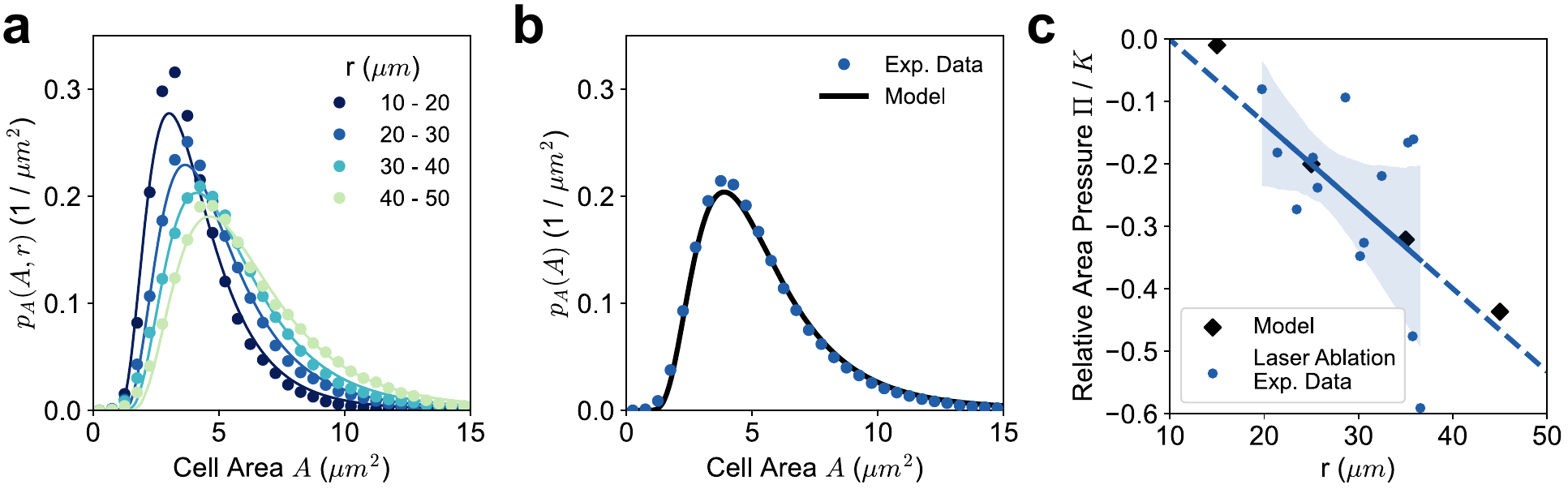}
    \caption{Spatial patterns of area are captured by varying pressure. (a) Probability distribution for cell area at different radii. Dots show experimental data. Lines show model predictions. (b) Model probability for cell area across the whole wing disc. Dots show experimental data. The line shows model predictions.
    (c) Relative area pressure $\Pi / K$ against radius from model predictions and laser ablation experiments from Ref. \cite{dye2021self}. Blue dots show the experimental data and the blue line is linear fit to the data. Shaded region corresponds to the uncertainty of the linear fit, and the dashed line is the extrapolation of the linear fit beyond the experimental range. Black dots show the model prediction. }
    \label{fig:4}
\end{figure*}

Until now we focused on normalised cell area. Now, we return to the description of actual cell areas. We first obtain the distribution of cell areas in each radial bin $p_A(A, r)= p(a) \overline{A}(r)$, which accounts for the experimentally observed cell area distributions as shown in Fig.~\ref{fig:4}(a). By summing the contributions from each radial bin weighted by the  number of cells in that bin $n(r)$, we find the prediction for the overall cell area distribution in the wing disc pouch
\begin{equation}
    p_{A}(A) = \frac{\sum_r p_A(A, r) n(r)}{\sum_r n(r)} .
\end{equation}
We find that our prediction achieves $R^2$-accuracy $99.3 \%$ when compared to the experimental data, see Fig.~\ref{fig:4}(b).

Since the cell area distributions in different radial bins differ only in the average cell area, we propose that this difference in $\overline{A}$ stems from the radial gradient of the average two-dimensional pressure $\Pi_0$. Measurement of the pressure, normalised by the area elastic modulus, in the wing disc pouch using laser ablation experiments was previously reported in Ref. \cite{dye2021self}. 
We use the pressure constitutive equation that we employ for pressure fluctuations also to describe spatial pressure variation. Therefore, we predict that pressure in each bin, normalised by the area elastic modulus $K$, is proportional to the logarithmic area strain $\Pi_0 (r)/K  = \ln{(\overline{A}(r) / A_p)}$, where $\overline{A}(r)$ is the average area in the bin and $A_p$ is a reference average area value attained at $0$ pressure. 
To test if this prediction is consistent with the laser ablation measurements from Ref. \cite{dye2021self}, we reproduce the laser ablation data in Fig.~\ref{fig:4}(c) and in addition we show the linear function fit to that data with the corresponding uncertainty interval. Our model predicts only changes of the pressure as a function of cell area, but not the absolute value of pressure. 
Therefore, to compare our prediction of pressure gradient with experimental data we adjust the reference area $A_p$ value to best match the laser ablation data. 
However, the prediction of the normalised pressure gradient is predicted by the model without fitting parameters, and we find that it is in good agreement with the laser ablation data. This confirms our hypothesis that spatial variations of the average cell area can be understood to stem from the radial pressure gradient.



\section*{Discussion}
Growth and division of cells naturally lead to the heterogeneity in cell areas in growing epithelial tissues. In this work, we studied the dynamics and statistics of cell areas in the pouch of a growing wing disc of the fruit fly \textit{D. melanogaster}. We found that the cell area distribution in different regions of the wing disc pouch are universal, up to scaling by the average value of the cell area. Furthermore, we discovered that a simple model of the cell cycle can account for about $85\%$ of the cell area distribution width. The remaining $15\%$ is accounted for by mechanical fluctuations. Importantly, we measure parameters of the cell cycle model, as well as the magnitude of the mechanical fluctuations, directly from the experimental data. In this way, we are able to compare the simple model of cell cycle and mechanical fluctuations to the experimental data without any fitting parameters and find that it fully accounts for the experimentally observed cell area distribution.

Our approach relies on extracting the growth rate, division rate and cell area fluctuations directly from the experimental data. In particular, on the coarse-grained level of our analysis, we find a strong dependence of these quantities on the target cell area. While the origin of these dependencies was not important to understand cell area distributions, it will be an interesting challenge to reproduce them from more detailed models of cell cycle dynamics. We expect the effective dependence of cell growth rate and division rate on the target cell area reported here will help constrain future models of cell cycle. 

Our results show that a growing tissue, undergoing growth and divisions, is expected to maintain a relatively high degree of cell area polydispersity of about $45 \%$, defined as the coefficient of variation of the distribution $P(a_0)$.
The fact that cell areas are likely heterogeneous in growing tissues is, however, typically not considered in models of tissue mechanics, where cells are often considered to be identical, and only recent work started to explicitly account for the cell cycle and divisions \cite{Bocanegra-Moreno2023}. The width of the cell area distributions, quantified by the area polydispersity, is a type of disorder in an epithelial tissue that can qualitatively affect its physical properties. The role of particle size polydispersity has been studied in systems of repulsive particles, where it can control the melting transition \cite{Sadr-Lahijany1997}, induce solid-solid phase separation \cite{Fasolo2003}. In athermal systems it can induce a transition from a crystal to an amorphous solid \cite{Tong2015}. Most recently, cell size polydispersity has been shown to control tissue crystallization \cite{Chhajed2025}. Therefore, quantification and understanding of the mechanisms that set the cell area polydispersity in biological tissues is crucial for development of realistic models of tissue mechanics.

\section*{Acknowledgments}
MP and NAD acknowledge funding by the Deutsche Forschungsgemeinschaft (DFG, German Research Foundation) - Project number 544201605, DY 180/2-1 and PO 3023/2-1. NAD additionally acknowledges funding from Deutsche Krebshilfe/MSNZ Dresden.

\section*{Methods}

\subsection*{Area statistics}


To measure the variance in area of a single cell, indexed by $c$, we record the variance of the normalised cell area within an hour window about its mean
\begin{equation}
    S^2_c(t) = \frac{1}{n}\sum_{k = 0}^{n} (a_c(t + k \Delta t) - a_{c,0}(t))^2 ,
\end{equation}
where
\begin{equation}
    a_{c,0}(t) = \frac{1}{n}\sum_{k=0}^{n} a(t + k \Delta t) 
\end{equation}
is the average cell area within the window, with $n = 12$ being the number of frames per hour. Next, we bin all data points by the one hour moving average normalised area to obtain a mean variance for given normalised cell area (Fig.~\ref{fig:2}(c)).

To estimate the division rate, we flag cells as beginning the division process one hour before they divide. This gives the latest moving average area that is not influenced by the large change in cell area during division, as the division process takes around 30 minutes to complete and we measure the moving average over an hour window. Next, we count the proportion of cells about to divide compared to the total number of cells at a given area, and divide by the duration of a time-frame to obtain a division rate estimate (Fig.~\ref{fig:2}(d)).

\subsection*{Numerical solution of the target cell area distribution equation}

To numerically solve the target area distribution in Eq.~\ref{eq:p0} we employ a finite difference scheme solving the time-dependent equation
\begin{align}
    \begin{split}
   \partial_t p_0(a_0)= &- \partial_{a_0}(\gamma(a_0) a_0 p_0(a_0))\\ &+ 2 \omega(2 a_0) p_0(2 a_0) \\ &- \omega(a_0)p_0(a_0) - \lambda p_0(a_0)
    \end{split}
\end{align}
until a steady state is reached.


\subsection*{Conditional area distribution}

The probability density of the pressure fluctuation, $p_\Pi(\Pi, t)$ obeys the Fokker-Planck equation
\begin{equation}
     D \partial^2_\Pi p_\Pi = \frac{1}{\tau}\partial_\Pi (\Pi p_\Pi)
\end{equation}
where $D$ is the variance of the Gaussian white noise. The left hand side represents the diffusion of probability due to stochastic noise, while the right hand side represents the relaxation of the pressure fluctuation.

Integrating once with respect to $\Pi$, and noting that the integration constant is zero since we have no probability at infinite pressure, we have the equation
\begin{equation}
    0 = D \partial_\Pi p_\Pi + \frac{1}{\tau}\Pi p_\Pi
\end{equation}
which has the solution 
\begin{equation}
    p_\Pi(\Pi) = \frac{1}{\sqrt{2 \pi D \tau}} e^{-\frac{1}{2 D \tau} \Pi^2}.
\end{equation}
This is the probability density of the normal distribution with mean $0$ and variance $D \tau$.

To obtain the probability density of cell area, first consider that since pressure is a monotonic function of area, the probability that our normalised area is less than some value, $P(a < a')$, is the same as the probability that the pressure is less than the corresponding pressure $P(\Pi(a) < \Pi(a'))$. Writing in integral form and substituting variables we obtain
\begin{equation}
    \int_{\Pi(0)}^{\Pi(a')} p_\Pi(\Pi) d\Pi = \int_0^{a'} p_\Pi(a) \frac{d\Pi}{da}da
\end{equation}
thus our area probability density is given by
\begin{equation}
    p(a | a_0) = p_\Pi(\Pi(a, a_0), \Pi_0) \frac{d\Pi}{da}.
\end{equation}
Substituting the normal distribution density function we obtain
\begin{equation}
    \label{SI:eq:p|a0}
    p(a | a_0) = \frac{1}{Z}e^{-\frac{1}{2D\tau} \left(\Pi(a, a_0) - \Pi_0\right)^2} \frac{d\Pi}{da}
\end{equation}
where $Z = \int_{-\infty}^{\Pi(0)} e^{-\frac{k}{D} \left(\Pi(a, a_0) - \Pi_0\right)^2}$ is the normalisation constant.

\bibliographystyle{apsrev4-1}
\bibliography{bib}

\begin{thebibliography}{36}%
\makeatletter
\providecommand \@ifxundefined [1]{%
 \@ifx{#1\undefined}
}%
\providecommand \@ifnum [1]{%
 \ifnum #1\expandafter \@firstoftwo
 \else \expandafter \@secondoftwo
 \fi
}%
\providecommand \@ifx [1]{%
 \ifx #1\expandafter \@firstoftwo
 \else \expandafter \@secondoftwo
 \fi
}%
\providecommand \natexlab [1]{#1}%
\providecommand \enquote  [1]{``#1''}%
\providecommand \bibnamefont  [1]{#1}%
\providecommand \bibfnamefont [1]{#1}%
\providecommand \citenamefont [1]{#1}%
\providecommand \href@noop [0]{\@secondoftwo}%
\providecommand \href [0]{\begingroup \@sanitize@url \@href}%
\providecommand \@href[1]{\@@startlink{#1}\@@href}%
\providecommand \@@href[1]{\endgroup#1\@@endlink}%
\providecommand \@sanitize@url [0]{\catcode `\\12\catcode `\$12\catcode `\&12\catcode `\#12\catcode `\^12\catcode `\_12\catcode `\%12\relax}%
\providecommand \@@startlink[1]{}%
\providecommand \@@endlink[0]{}%
\providecommand \url  [0]{\begingroup\@sanitize@url \@url }%
\providecommand \@url [1]{\endgroup\@href {#1}{\urlprefix }}%
\providecommand \urlprefix  [0]{URL }%
\providecommand \Eprint [0]{\href }%
\providecommand \doibase [0]{http://dx.doi.org/}%
\providecommand \selectlanguage [0]{\@gobble}%
\providecommand \bibinfo  [0]{\@secondoftwo}%
\providecommand \bibfield  [0]{\@secondoftwo}%
\providecommand \translation [1]{[#1]}%
\providecommand \BibitemOpen [0]{}%
\providecommand \bibitemStop [0]{}%
\providecommand \bibitemNoStop [0]{.\EOS\space}%
\providecommand \EOS [0]{\spacefactor3000\relax}%
\providecommand \BibitemShut  [1]{\csname bibitem#1\endcsname}%
\let\auto@bib@innerbib\@empty
\bibitem [{\citenamefont {Friedl}\ and\ \citenamefont {Gilmour}(2009)}]{friedl2009collective}%
  \BibitemOpen
  \bibfield  {author} {\bibinfo {author} {\bibfnamefont {P.}~\bibnamefont {Friedl}}\ and\ \bibinfo {author} {\bibfnamefont {D.}~\bibnamefont {Gilmour}},\ }\href@noop {} {\bibfield  {journal} {\bibinfo  {journal} {Nature reviews Molecular cell biology}\ }\textbf {\bibinfo {volume} {10}},\ \bibinfo {pages} {445} (\bibinfo {year} {2009})}\BibitemShut {NoStop}%
\bibitem [{\citenamefont {Behrndt}\ \emph {et~al.}(2012)\citenamefont {Behrndt}, \citenamefont {Salbreux}, \citenamefont {Campinho}, \citenamefont {Hauschild}, \citenamefont {Oswald}, \citenamefont {Roensch}, \citenamefont {Grill},\ and\ \citenamefont {Heisenberg}}]{behrndt2012forces}%
  \BibitemOpen
  \bibfield  {author} {\bibinfo {author} {\bibfnamefont {M.}~\bibnamefont {Behrndt}}, \bibinfo {author} {\bibfnamefont {G.}~\bibnamefont {Salbreux}}, \bibinfo {author} {\bibfnamefont {P.}~\bibnamefont {Campinho}}, \bibinfo {author} {\bibfnamefont {R.}~\bibnamefont {Hauschild}}, \bibinfo {author} {\bibfnamefont {F.}~\bibnamefont {Oswald}}, \bibinfo {author} {\bibfnamefont {J.}~\bibnamefont {Roensch}}, \bibinfo {author} {\bibfnamefont {S.~W.}\ \bibnamefont {Grill}}, \ and\ \bibinfo {author} {\bibfnamefont {C.-P.}\ \bibnamefont {Heisenberg}},\ }\href@noop {} {\bibfield  {journal} {\bibinfo  {journal} {Science}\ }\textbf {\bibinfo {volume} {338}},\ \bibinfo {pages} {257} (\bibinfo {year} {2012})}\BibitemShut {NoStop}%
\bibitem [{\citenamefont {Etournay}\ \emph {et~al.}(2015)\citenamefont {Etournay}, \citenamefont {Popovi{\'c}}, \citenamefont {Merkel}, \citenamefont {Nandi}, \citenamefont {Blasse}, \citenamefont {Aigouy}, \citenamefont {Brandl}, \citenamefont {Myers}, \citenamefont {Salbreux}, \citenamefont {J{\"u}licher} \emph {et~al.}}]{etournay2015interplay}%
  \BibitemOpen
  \bibfield  {author} {\bibinfo {author} {\bibfnamefont {R.}~\bibnamefont {Etournay}}, \bibinfo {author} {\bibfnamefont {M.}~\bibnamefont {Popovi{\'c}}}, \bibinfo {author} {\bibfnamefont {M.}~\bibnamefont {Merkel}}, \bibinfo {author} {\bibfnamefont {A.}~\bibnamefont {Nandi}}, \bibinfo {author} {\bibfnamefont {C.}~\bibnamefont {Blasse}}, \bibinfo {author} {\bibfnamefont {B.}~\bibnamefont {Aigouy}}, \bibinfo {author} {\bibfnamefont {H.}~\bibnamefont {Brandl}}, \bibinfo {author} {\bibfnamefont {G.}~\bibnamefont {Myers}}, \bibinfo {author} {\bibfnamefont {G.}~\bibnamefont {Salbreux}}, \bibinfo {author} {\bibfnamefont {F.}~\bibnamefont {J{\"u}licher}},  \emph {et~al.},\ }\href@noop {} {\bibfield  {journal} {\bibinfo  {journal} {Elife}\ }\textbf {\bibinfo {volume} {4}},\ \bibinfo {pages} {e07090} (\bibinfo {year} {2015})}\BibitemShut {NoStop}%
\bibitem [{\citenamefont {J{\"u}licher}\ and\ \citenamefont {Eaton}(2017)}]{julicher2017emergence}%
  \BibitemOpen
  \bibfield  {author} {\bibinfo {author} {\bibfnamefont {F.}~\bibnamefont {J{\"u}licher}}\ and\ \bibinfo {author} {\bibfnamefont {S.}~\bibnamefont {Eaton}},\ }in\ \href@noop {} {\emph {\bibinfo {booktitle} {Seminars in cell \& developmental biology}}},\ Vol.~\bibinfo {volume} {67}\ (\bibinfo {organization} {Elsevier},\ \bibinfo {year} {2017})\ pp.\ \bibinfo {pages} {103--112}\BibitemShut {NoStop}%
\bibitem [{\citenamefont {Maniou}\ \emph {et~al.}(2021)\citenamefont {Maniou}, \citenamefont {Staddon}, \citenamefont {Marshall}, \citenamefont {Greene}, \citenamefont {Copp}, \citenamefont {Banerjee},\ and\ \citenamefont {Galea}}]{maniou2021hindbrain}%
  \BibitemOpen
  \bibfield  {author} {\bibinfo {author} {\bibfnamefont {E.}~\bibnamefont {Maniou}}, \bibinfo {author} {\bibfnamefont {M.~F.}\ \bibnamefont {Staddon}}, \bibinfo {author} {\bibfnamefont {A.~R.}\ \bibnamefont {Marshall}}, \bibinfo {author} {\bibfnamefont {N.~D.}\ \bibnamefont {Greene}}, \bibinfo {author} {\bibfnamefont {A.~J.}\ \bibnamefont {Copp}}, \bibinfo {author} {\bibfnamefont {S.}~\bibnamefont {Banerjee}}, \ and\ \bibinfo {author} {\bibfnamefont {G.~L.}\ \bibnamefont {Galea}},\ }\href@noop {} {\bibfield  {journal} {\bibinfo  {journal} {Proceedings of the National Academy of Sciences}\ }\textbf {\bibinfo {volume} {118}},\ \bibinfo {pages} {e2023163118} (\bibinfo {year} {2021})}\BibitemShut {NoStop}%
\bibitem [{\citenamefont {Friedl}\ \emph {et~al.}(2012)\citenamefont {Friedl}, \citenamefont {Locker}, \citenamefont {Sahai},\ and\ \citenamefont {Segall}}]{friedl2012classifying}%
  \BibitemOpen
  \bibfield  {author} {\bibinfo {author} {\bibfnamefont {P.}~\bibnamefont {Friedl}}, \bibinfo {author} {\bibfnamefont {J.}~\bibnamefont {Locker}}, \bibinfo {author} {\bibfnamefont {E.}~\bibnamefont {Sahai}}, \ and\ \bibinfo {author} {\bibfnamefont {J.~E.}\ \bibnamefont {Segall}},\ }\href@noop {} {\bibfield  {journal} {\bibinfo  {journal} {Nature cell biology}\ }\textbf {\bibinfo {volume} {14}},\ \bibinfo {pages} {777} (\bibinfo {year} {2012})}\BibitemShut {NoStop}%
\bibitem [{\citenamefont {Arwert}\ \emph {et~al.}(2012)\citenamefont {Arwert}, \citenamefont {Hoste},\ and\ \citenamefont {Watt}}]{arwert2012epithelial}%
  \BibitemOpen
  \bibfield  {author} {\bibinfo {author} {\bibfnamefont {E.~N.}\ \bibnamefont {Arwert}}, \bibinfo {author} {\bibfnamefont {E.}~\bibnamefont {Hoste}}, \ and\ \bibinfo {author} {\bibfnamefont {F.~M.}\ \bibnamefont {Watt}},\ }\href@noop {} {\bibfield  {journal} {\bibinfo  {journal} {Nature Reviews Cancer}\ }\textbf {\bibinfo {volume} {12}},\ \bibinfo {pages} {170} (\bibinfo {year} {2012})}\BibitemShut {NoStop}%
\bibitem [{\citenamefont {Brugu{\'e}s}\ \emph {et~al.}(2014)\citenamefont {Brugu{\'e}s}, \citenamefont {Anon}, \citenamefont {Conte}, \citenamefont {Veldhuis}, \citenamefont {Gupta}, \citenamefont {Colombelli}, \citenamefont {Mu{\~n}oz}, \citenamefont {Brodland}, \citenamefont {Ladoux},\ and\ \citenamefont {Trepat}}]{brugues2014forces}%
  \BibitemOpen
  \bibfield  {author} {\bibinfo {author} {\bibfnamefont {A.}~\bibnamefont {Brugu{\'e}s}}, \bibinfo {author} {\bibfnamefont {E.}~\bibnamefont {Anon}}, \bibinfo {author} {\bibfnamefont {V.}~\bibnamefont {Conte}}, \bibinfo {author} {\bibfnamefont {J.~H.}\ \bibnamefont {Veldhuis}}, \bibinfo {author} {\bibfnamefont {M.}~\bibnamefont {Gupta}}, \bibinfo {author} {\bibfnamefont {J.}~\bibnamefont {Colombelli}}, \bibinfo {author} {\bibfnamefont {J.~J.}\ \bibnamefont {Mu{\~n}oz}}, \bibinfo {author} {\bibfnamefont {G.~W.}\ \bibnamefont {Brodland}}, \bibinfo {author} {\bibfnamefont {B.}~\bibnamefont {Ladoux}}, \ and\ \bibinfo {author} {\bibfnamefont {X.}~\bibnamefont {Trepat}},\ }\href@noop {} {\bibfield  {journal} {\bibinfo  {journal} {Nature physics}\ }\textbf {\bibinfo {volume} {10}},\ \bibinfo {pages} {683} (\bibinfo {year} {2014})}\BibitemShut {NoStop}%
\bibitem [{\citenamefont {Tetley}\ \emph {et~al.}(2019)\citenamefont {Tetley}, \citenamefont {Staddon}, \citenamefont {Heller}, \citenamefont {Hoppe}, \citenamefont {Banerjee},\ and\ \citenamefont {Mao}}]{tetley2019tissue}%
  \BibitemOpen
  \bibfield  {author} {\bibinfo {author} {\bibfnamefont {R.~J.}\ \bibnamefont {Tetley}}, \bibinfo {author} {\bibfnamefont {M.~F.}\ \bibnamefont {Staddon}}, \bibinfo {author} {\bibfnamefont {D.}~\bibnamefont {Heller}}, \bibinfo {author} {\bibfnamefont {A.}~\bibnamefont {Hoppe}}, \bibinfo {author} {\bibfnamefont {S.}~\bibnamefont {Banerjee}}, \ and\ \bibinfo {author} {\bibfnamefont {Y.}~\bibnamefont {Mao}},\ }\href@noop {} {\bibfield  {journal} {\bibinfo  {journal} {Nature physics}\ }\textbf {\bibinfo {volume} {15}},\ \bibinfo {pages} {1195} (\bibinfo {year} {2019})}\BibitemShut {NoStop}%
\bibitem [{\citenamefont {Ajeti}\ \emph {et~al.}(2019)\citenamefont {Ajeti}, \citenamefont {Tabatabai}, \citenamefont {Fleszar}, \citenamefont {Staddon}, \citenamefont {Seara}, \citenamefont {Suarez}, \citenamefont {Yousafzai}, \citenamefont {Bi}, \citenamefont {Kovar}, \citenamefont {Banerjee} \emph {et~al.}}]{ajeti2019wound}%
  \BibitemOpen
  \bibfield  {author} {\bibinfo {author} {\bibfnamefont {V.}~\bibnamefont {Ajeti}}, \bibinfo {author} {\bibfnamefont {A.~P.}\ \bibnamefont {Tabatabai}}, \bibinfo {author} {\bibfnamefont {A.~J.}\ \bibnamefont {Fleszar}}, \bibinfo {author} {\bibfnamefont {M.~F.}\ \bibnamefont {Staddon}}, \bibinfo {author} {\bibfnamefont {D.~S.}\ \bibnamefont {Seara}}, \bibinfo {author} {\bibfnamefont {C.}~\bibnamefont {Suarez}}, \bibinfo {author} {\bibfnamefont {M.~S.}\ \bibnamefont {Yousafzai}}, \bibinfo {author} {\bibfnamefont {D.}~\bibnamefont {Bi}}, \bibinfo {author} {\bibfnamefont {D.~R.}\ \bibnamefont {Kovar}}, \bibinfo {author} {\bibfnamefont {S.}~\bibnamefont {Banerjee}},  \emph {et~al.},\ }\href@noop {} {\bibfield  {journal} {\bibinfo  {journal} {Nature physics}\ }\textbf {\bibinfo {volume} {15}},\ \bibinfo {pages} {696} (\bibinfo {year} {2019})}\BibitemShut {NoStop}%
\bibitem [{\citenamefont {Salbreux}\ \emph {et~al.}(2012)\citenamefont {Salbreux}, \citenamefont {Charras},\ and\ \citenamefont {Paluch}}]{salbreux2012actin}%
  \BibitemOpen
  \bibfield  {author} {\bibinfo {author} {\bibfnamefont {G.}~\bibnamefont {Salbreux}}, \bibinfo {author} {\bibfnamefont {G.}~\bibnamefont {Charras}}, \ and\ \bibinfo {author} {\bibfnamefont {E.}~\bibnamefont {Paluch}},\ }\href@noop {} {\bibfield  {journal} {\bibinfo  {journal} {Trends in cell biology}\ }\textbf {\bibinfo {volume} {22}},\ \bibinfo {pages} {536} (\bibinfo {year} {2012})}\BibitemShut {NoStop}%
\bibitem [{\citenamefont {Murrell}\ \emph {et~al.}(2015)\citenamefont {Murrell}, \citenamefont {Oakes}, \citenamefont {Lenz},\ and\ \citenamefont {Gardel}}]{murrell2015forcing}%
  \BibitemOpen
  \bibfield  {author} {\bibinfo {author} {\bibfnamefont {M.}~\bibnamefont {Murrell}}, \bibinfo {author} {\bibfnamefont {P.~W.}\ \bibnamefont {Oakes}}, \bibinfo {author} {\bibfnamefont {M.}~\bibnamefont {Lenz}}, \ and\ \bibinfo {author} {\bibfnamefont {M.~L.}\ \bibnamefont {Gardel}},\ }\href@noop {} {\bibfield  {journal} {\bibinfo  {journal} {Nature reviews Molecular cell biology}\ }\textbf {\bibinfo {volume} {16}},\ \bibinfo {pages} {486} (\bibinfo {year} {2015})}\BibitemShut {NoStop}%
\bibitem [{\citenamefont {Ladoux}\ and\ \citenamefont {M{\`e}ge}(2017)}]{ladoux2017mechanobiology}%
  \BibitemOpen
  \bibfield  {author} {\bibinfo {author} {\bibfnamefont {B.}~\bibnamefont {Ladoux}}\ and\ \bibinfo {author} {\bibfnamefont {R.-M.}\ \bibnamefont {M{\`e}ge}},\ }\href@noop {} {\bibfield  {journal} {\bibinfo  {journal} {Nature reviews Molecular cell biology}\ }\textbf {\bibinfo {volume} {18}},\ \bibinfo {pages} {743} (\bibinfo {year} {2017})}\BibitemShut {NoStop}%
\bibitem [{\citenamefont {Firmino}\ \emph {et~al.}(2016)\citenamefont {Firmino}, \citenamefont {Rocancourt}, \citenamefont {Saadaoui}, \citenamefont {Moreau},\ and\ \citenamefont {Gros}}]{firmino2016cell}%
  \BibitemOpen
  \bibfield  {author} {\bibinfo {author} {\bibfnamefont {J.}~\bibnamefont {Firmino}}, \bibinfo {author} {\bibfnamefont {D.}~\bibnamefont {Rocancourt}}, \bibinfo {author} {\bibfnamefont {M.}~\bibnamefont {Saadaoui}}, \bibinfo {author} {\bibfnamefont {C.}~\bibnamefont {Moreau}}, \ and\ \bibinfo {author} {\bibfnamefont {J.}~\bibnamefont {Gros}},\ }\href@noop {} {\bibfield  {journal} {\bibinfo  {journal} {Developmental cell}\ }\textbf {\bibinfo {volume} {36}},\ \bibinfo {pages} {249} (\bibinfo {year} {2016})}\BibitemShut {NoStop}%
\bibitem [{\citenamefont {Devany}\ \emph {et~al.}(2021)\citenamefont {Devany}, \citenamefont {Sussman}, \citenamefont {Yamamoto}, \citenamefont {Manning},\ and\ \citenamefont {Gardel}}]{devany2021cell}%
  \BibitemOpen
  \bibfield  {author} {\bibinfo {author} {\bibfnamefont {J.}~\bibnamefont {Devany}}, \bibinfo {author} {\bibfnamefont {D.~M.}\ \bibnamefont {Sussman}}, \bibinfo {author} {\bibfnamefont {T.}~\bibnamefont {Yamamoto}}, \bibinfo {author} {\bibfnamefont {M.~L.}\ \bibnamefont {Manning}}, \ and\ \bibinfo {author} {\bibfnamefont {M.~L.}\ \bibnamefont {Gardel}},\ }\href@noop {} {\bibfield  {journal} {\bibinfo  {journal} {Proceedings of the National Academy of Sciences}\ }\textbf {\bibinfo {volume} {118}},\ \bibinfo {pages} {e1917853118} (\bibinfo {year} {2021})}\BibitemShut {NoStop}%
\bibitem [{\citenamefont {Tahaei}\ \emph {et~al.}(2024)\citenamefont {Tahaei}, \citenamefont {Pisticello-G{\' o}mez}, \citenamefont {Suganthan}, \citenamefont {Cwikla}, \citenamefont {Fuhrmann}, \citenamefont {Dye},\ and\ \citenamefont {Popovi{\' c}}}]{Tahaei2024}%
  \BibitemOpen
  \bibfield  {author} {\bibinfo {author} {\bibfnamefont {A.}~\bibnamefont {Tahaei}}, \bibinfo {author} {\bibfnamefont {R.}~\bibnamefont {Pisticello-G{\' o}mez}}, \bibinfo {author} {\bibfnamefont {S.}~\bibnamefont {Suganthan}}, \bibinfo {author} {\bibfnamefont {G.}~\bibnamefont {Cwikla}}, \bibinfo {author} {\bibfnamefont {J.~F.}\ \bibnamefont {Fuhrmann}}, \bibinfo {author} {\bibfnamefont {N.~A.}\ \bibnamefont {Dye}}, \ and\ \bibinfo {author} {\bibfnamefont {M.}~\bibnamefont {Popovi{\' c}}},\ }\href {http://arxiv.org/abs/2406.03433} {\  (\bibinfo {year} {2024})},\ \bibinfo {note} {arXiv:2406.03433 [cond-mat, physics:physics]}\BibitemShut {NoStop}%
\bibitem [{\citenamefont {Ranft}\ \emph {et~al.}(2010)\citenamefont {Ranft}, \citenamefont {Basan}, \citenamefont {Elgeti}, \citenamefont {Joanny}, \citenamefont {Prost},\ and\ \citenamefont {J{\"u}licher}}]{ranft2010fluidization}%
  \BibitemOpen
  \bibfield  {author} {\bibinfo {author} {\bibfnamefont {J.}~\bibnamefont {Ranft}}, \bibinfo {author} {\bibfnamefont {M.}~\bibnamefont {Basan}}, \bibinfo {author} {\bibfnamefont {J.}~\bibnamefont {Elgeti}}, \bibinfo {author} {\bibfnamefont {J.-F.}\ \bibnamefont {Joanny}}, \bibinfo {author} {\bibfnamefont {J.}~\bibnamefont {Prost}}, \ and\ \bibinfo {author} {\bibfnamefont {F.}~\bibnamefont {J{\"u}licher}},\ }\href@noop {} {\bibfield  {journal} {\bibinfo  {journal} {Proceedings of the National Academy of Sciences}\ }\textbf {\bibinfo {volume} {107}},\ \bibinfo {pages} {20863} (\bibinfo {year} {2010})}\BibitemShut {NoStop}%
\bibitem [{\citenamefont {Godard}\ and\ \citenamefont {Heisenberg}(2019)}]{godard2019cell}%
  \BibitemOpen
  \bibfield  {author} {\bibinfo {author} {\bibfnamefont {B.~G.}\ \bibnamefont {Godard}}\ and\ \bibinfo {author} {\bibfnamefont {C.-P.}\ \bibnamefont {Heisenberg}},\ }\href@noop {} {\bibfield  {journal} {\bibinfo  {journal} {Current opinion in cell biology}\ }\textbf {\bibinfo {volume} {60}},\ \bibinfo {pages} {114} (\bibinfo {year} {2019})}\BibitemShut {NoStop}%
\bibitem [{\citenamefont {Bocanegra-Moreno}\ \emph {et~al.}(2022)\citenamefont {Bocanegra-Moreno}, \citenamefont {Singh}, \citenamefont {Hannezo}, \citenamefont {Zagorski},\ and\ \citenamefont {Kicheva}}]{bocanegra2022cell}%
  \BibitemOpen
  \bibfield  {author} {\bibinfo {author} {\bibfnamefont {L.}~\bibnamefont {Bocanegra-Moreno}}, \bibinfo {author} {\bibfnamefont {A.}~\bibnamefont {Singh}}, \bibinfo {author} {\bibfnamefont {E.}~\bibnamefont {Hannezo}}, \bibinfo {author} {\bibfnamefont {M.}~\bibnamefont {Zagorski}}, \ and\ \bibinfo {author} {\bibfnamefont {A.}~\bibnamefont {Kicheva}},\ }\href@noop {} {\bibfield  {journal} {\bibinfo  {journal} {BioRxiv}\ } (\bibinfo {year} {2022})}\BibitemShut {NoStop}%
\bibitem [{\citenamefont {Wyatt}\ \emph {et~al.}(2015)\citenamefont {Wyatt}, \citenamefont {Harris}, \citenamefont {Lam}, \citenamefont {Cheng}, \citenamefont {Bellis}, \citenamefont {Dimitracopoulos}, \citenamefont {Kabla}, \citenamefont {Charras},\ and\ \citenamefont {Baum}}]{wyatt2015emergence}%
  \BibitemOpen
  \bibfield  {author} {\bibinfo {author} {\bibfnamefont {T.~P.}\ \bibnamefont {Wyatt}}, \bibinfo {author} {\bibfnamefont {A.~R.}\ \bibnamefont {Harris}}, \bibinfo {author} {\bibfnamefont {M.}~\bibnamefont {Lam}}, \bibinfo {author} {\bibfnamefont {Q.}~\bibnamefont {Cheng}}, \bibinfo {author} {\bibfnamefont {J.}~\bibnamefont {Bellis}}, \bibinfo {author} {\bibfnamefont {A.}~\bibnamefont {Dimitracopoulos}}, \bibinfo {author} {\bibfnamefont {A.~J.}\ \bibnamefont {Kabla}}, \bibinfo {author} {\bibfnamefont {G.~T.}\ \bibnamefont {Charras}}, \ and\ \bibinfo {author} {\bibfnamefont {B.}~\bibnamefont {Baum}},\ }\href@noop {} {\bibfield  {journal} {\bibinfo  {journal} {Proceedings of the National Academy of Sciences}\ }\textbf {\bibinfo {volume} {112}},\ \bibinfo {pages} {5726} (\bibinfo {year} {2015})}\BibitemShut {NoStop}%
\bibitem [{\citenamefont {Hirano}\ \emph {et~al.}(1982)\citenamefont {Hirano}, \citenamefont {Nordheim}, \citenamefont {Arny},\ and\ \citenamefont {Upper}}]{hirano1982lognormal}%
  \BibitemOpen
  \bibfield  {author} {\bibinfo {author} {\bibfnamefont {S.~S.}\ \bibnamefont {Hirano}}, \bibinfo {author} {\bibfnamefont {E.~V.}\ \bibnamefont {Nordheim}}, \bibinfo {author} {\bibfnamefont {D.~C.}\ \bibnamefont {Arny}}, \ and\ \bibinfo {author} {\bibfnamefont {C.~D.}\ \bibnamefont {Upper}},\ }\href@noop {} {\bibfield  {journal} {\bibinfo  {journal} {Applied and Environmental Microbiology}\ }\textbf {\bibinfo {volume} {44}},\ \bibinfo {pages} {695} (\bibinfo {year} {1982})}\BibitemShut {NoStop}%
\bibitem [{\citenamefont {Loper}\ \emph {et~al.}(1984)\citenamefont {Loper}, \citenamefont {Suslow},\ and\ \citenamefont {Schroth}}]{loper1984lognormal}%
  \BibitemOpen
  \bibfield  {author} {\bibinfo {author} {\bibfnamefont {J.}~\bibnamefont {Loper}}, \bibinfo {author} {\bibfnamefont {T.}~\bibnamefont {Suslow}}, \ and\ \bibinfo {author} {\bibfnamefont {M.}~\bibnamefont {Schroth}},\ }\href@noop {} {\bibfield  {journal} {\bibinfo  {journal} {Phytopathology}\ }\textbf {\bibinfo {volume} {74}},\ \bibinfo {pages} {1454} (\bibinfo {year} {1984})}\BibitemShut {NoStop}%
\bibitem [{\citenamefont {Koutsoumanis}\ and\ \citenamefont {Lianou}(2013)}]{koutsoumanis2013stochasticity}%
  \BibitemOpen
  \bibfield  {author} {\bibinfo {author} {\bibfnamefont {K.~P.}\ \bibnamefont {Koutsoumanis}}\ and\ \bibinfo {author} {\bibfnamefont {A.}~\bibnamefont {Lianou}},\ }\href@noop {} {\bibfield  {journal} {\bibinfo  {journal} {Applied and Environmental Microbiology}\ }\textbf {\bibinfo {volume} {79}},\ \bibinfo {pages} {2294} (\bibinfo {year} {2013})}\BibitemShut {NoStop}%
\bibitem [{\citenamefont {Zehnder}\ \emph {et~al.}(2015)\citenamefont {Zehnder}, \citenamefont {Suaris}, \citenamefont {Bellaire},\ and\ \citenamefont {Angelini}}]{zehnder2015cell}%
  \BibitemOpen
  \bibfield  {author} {\bibinfo {author} {\bibfnamefont {S.~M.}\ \bibnamefont {Zehnder}}, \bibinfo {author} {\bibfnamefont {M.}~\bibnamefont {Suaris}}, \bibinfo {author} {\bibfnamefont {M.~M.}\ \bibnamefont {Bellaire}}, \ and\ \bibinfo {author} {\bibfnamefont {T.~E.}\ \bibnamefont {Angelini}},\ }\href@noop {} {\bibfield  {journal} {\bibinfo  {journal} {Biophysical journal}\ }\textbf {\bibinfo {volume} {108}},\ \bibinfo {pages} {247} (\bibinfo {year} {2015})}\BibitemShut {NoStop}%
\bibitem [{\citenamefont {Puliafito}\ \emph {et~al.}(2017)\citenamefont {Puliafito}, \citenamefont {Primo},\ and\ \citenamefont {Celani}}]{puliafito2017cell}%
  \BibitemOpen
  \bibfield  {author} {\bibinfo {author} {\bibfnamefont {A.}~\bibnamefont {Puliafito}}, \bibinfo {author} {\bibfnamefont {L.}~\bibnamefont {Primo}}, \ and\ \bibinfo {author} {\bibfnamefont {A.}~\bibnamefont {Celani}},\ }\href@noop {} {\bibfield  {journal} {\bibinfo  {journal} {Journal of The Royal Society Interface}\ }\textbf {\bibinfo {volume} {14}},\ \bibinfo {pages} {20170032} (\bibinfo {year} {2017})}\BibitemShut {NoStop}%
\bibitem [{\citenamefont {Amir}(2014)}]{amir2014cell}%
  \BibitemOpen
  \bibfield  {author} {\bibinfo {author} {\bibfnamefont {A.}~\bibnamefont {Amir}},\ }\href@noop {} {\bibfield  {journal} {\bibinfo  {journal} {Physical review letters}\ }\textbf {\bibinfo {volume} {112}},\ \bibinfo {pages} {208102} (\bibinfo {year} {2014})}\BibitemShut {NoStop}%
\bibitem [{\citenamefont {Hosoda}\ \emph {et~al.}(2011)\citenamefont {Hosoda}, \citenamefont {Matsuura}, \citenamefont {Suzuki},\ and\ \citenamefont {Yomo}}]{hosoda2011origin}%
  \BibitemOpen
  \bibfield  {author} {\bibinfo {author} {\bibfnamefont {K.}~\bibnamefont {Hosoda}}, \bibinfo {author} {\bibfnamefont {T.}~\bibnamefont {Matsuura}}, \bibinfo {author} {\bibfnamefont {H.}~\bibnamefont {Suzuki}}, \ and\ \bibinfo {author} {\bibfnamefont {T.}~\bibnamefont {Yomo}},\ }\href@noop {} {\bibfield  {journal} {\bibinfo  {journal} {Physical Review E}\ }\textbf {\bibinfo {volume} {83}},\ \bibinfo {pages} {031118} (\bibinfo {year} {2011})}\BibitemShut {NoStop}%
\bibitem [{\citenamefont {Wilk}\ \emph {et~al.}(2014)\citenamefont {Wilk}, \citenamefont {Iwasa}, \citenamefont {Fuller}, \citenamefont {Kandere-Grzybowska},\ and\ \citenamefont {Grzybowski}}]{wilk2014universal}%
  \BibitemOpen
  \bibfield  {author} {\bibinfo {author} {\bibfnamefont {G.}~\bibnamefont {Wilk}}, \bibinfo {author} {\bibfnamefont {M.}~\bibnamefont {Iwasa}}, \bibinfo {author} {\bibfnamefont {P.~E.}\ \bibnamefont {Fuller}}, \bibinfo {author} {\bibfnamefont {K.}~\bibnamefont {Kandere-Grzybowska}}, \ and\ \bibinfo {author} {\bibfnamefont {B.~A.}\ \bibnamefont {Grzybowski}},\ }\href@noop {} {\bibfield  {journal} {\bibinfo  {journal} {Physical review letters}\ }\textbf {\bibinfo {volume} {112}},\ \bibinfo {pages} {138104} (\bibinfo {year} {2014})}\BibitemShut {NoStop}%
\bibitem [{\citenamefont {Genthon}(2022)}]{genthon2022analytical}%
  \BibitemOpen
  \bibfield  {author} {\bibinfo {author} {\bibfnamefont {A.}~\bibnamefont {Genthon}},\ }\href@noop {} {\bibfield  {journal} {\bibinfo  {journal} {Journal of the Royal Society Interface}\ }\textbf {\bibinfo {volume} {19}},\ \bibinfo {pages} {20220405} (\bibinfo {year} {2022})}\BibitemShut {NoStop}%
\bibitem [{\citenamefont {Dye}\ \emph {et~al.}(2021)\citenamefont {Dye}, \citenamefont {Popovi{\'c}}, \citenamefont {Iyer}, \citenamefont {Fuhrmann}, \citenamefont {Piscitello-G{\'o}mez}, \citenamefont {Eaton},\ and\ \citenamefont {J{\"u}licher}}]{dye2021self}%
  \BibitemOpen
  \bibfield  {author} {\bibinfo {author} {\bibfnamefont {N.~A.}\ \bibnamefont {Dye}}, \bibinfo {author} {\bibfnamefont {M.}~\bibnamefont {Popovi{\'c}}}, \bibinfo {author} {\bibfnamefont {K.~V.}\ \bibnamefont {Iyer}}, \bibinfo {author} {\bibfnamefont {J.~F.}\ \bibnamefont {Fuhrmann}}, \bibinfo {author} {\bibfnamefont {R.}~\bibnamefont {Piscitello-G{\'o}mez}}, \bibinfo {author} {\bibfnamefont {S.}~\bibnamefont {Eaton}}, \ and\ \bibinfo {author} {\bibfnamefont {F.}~\bibnamefont {J{\"u}licher}},\ }\href@noop {} {\bibfield  {journal} {\bibinfo  {journal} {Elife}\ }\textbf {\bibinfo {volume} {10}},\ \bibinfo {pages} {e57964} (\bibinfo {year} {2021})}\BibitemShut {NoStop}%
\bibitem [{\citenamefont {Buchmann}\ \emph {et~al.}(2014)\citenamefont {Buchmann}, \citenamefont {Alber},\ and\ \citenamefont {Zartman}}]{Buchmann2014}%
  \BibitemOpen
  \bibfield  {author} {\bibinfo {author} {\bibfnamefont {A.}~\bibnamefont {Buchmann}}, \bibinfo {author} {\bibfnamefont {M.}~\bibnamefont {Alber}}, \ and\ \bibinfo {author} {\bibfnamefont {J.~J.}\ \bibnamefont {Zartman}},\ }\href {\doibase 10.1016/j.semcdb.2014.06.018} {\bibfield  {journal} {\bibinfo  {journal} {Seminars in Cell \& Developmental Biology}\ }\textbf {\bibinfo {volume} {35}},\ \bibinfo {pages} {73–81} (\bibinfo {year} {2014})}\BibitemShut {NoStop}%
\bibitem [{\citenamefont {Bocanegra-Moreno}\ \emph {et~al.}(2023)\citenamefont {Bocanegra-Moreno}, \citenamefont {Singh}, \citenamefont {Hannezo}, \citenamefont {Zagorski},\ and\ \citenamefont {Kicheva}}]{Bocanegra-Moreno2023}%
  \BibitemOpen
  \bibfield  {author} {\bibinfo {author} {\bibfnamefont {L.}~\bibnamefont {Bocanegra-Moreno}}, \bibinfo {author} {\bibfnamefont {A.}~\bibnamefont {Singh}}, \bibinfo {author} {\bibfnamefont {E.}~\bibnamefont {Hannezo}}, \bibinfo {author} {\bibfnamefont {M.}~\bibnamefont {Zagorski}}, \ and\ \bibinfo {author} {\bibfnamefont {A.}~\bibnamefont {Kicheva}},\ }\href {\doibase 10.1038/s41567-023-01977-w} {\bibfield  {journal} {\bibinfo  {journal} {Nature Physics}\ }\textbf {\bibinfo {volume} {19}},\ \bibinfo {pages} {1050–1058} (\bibinfo {year} {2023})}\BibitemShut {NoStop}%
\bibitem [{\citenamefont {Sadr-Lahijany}\ \emph {et~al.}(1997)\citenamefont {Sadr-Lahijany}, \citenamefont {Ray},\ and\ \citenamefont {Stanley}}]{Sadr-Lahijany1997}%
  \BibitemOpen
  \bibfield  {author} {\bibinfo {author} {\bibfnamefont {M.~R.}\ \bibnamefont {Sadr-Lahijany}}, \bibinfo {author} {\bibfnamefont {P.}~\bibnamefont {Ray}}, \ and\ \bibinfo {author} {\bibfnamefont {H.~E.}\ \bibnamefont {Stanley}},\ }\href {\doibase 10.1103/PhysRevLett.79.3206} {\bibfield  {journal} {\bibinfo  {journal} {Physical Review Letters}\ }\textbf {\bibinfo {volume} {79}},\ \bibinfo {pages} {3206–3209} (\bibinfo {year} {1997})}\BibitemShut {NoStop}%
\bibitem [{\citenamefont {Fasolo}\ and\ \citenamefont {Sollich}(2003)}]{Fasolo2003}%
  \BibitemOpen
  \bibfield  {author} {\bibinfo {author} {\bibfnamefont {M.}~\bibnamefont {Fasolo}}\ and\ \bibinfo {author} {\bibfnamefont {P.}~\bibnamefont {Sollich}},\ }\href {\doibase 10.1103/PhysRevLett.91.068301} {\bibfield  {journal} {\bibinfo  {journal} {Physical Review Letters}\ }\textbf {\bibinfo {volume} {91}},\ \bibinfo {pages} {068301} (\bibinfo {year} {2003})}\BibitemShut {NoStop}%
\bibitem [{\citenamefont {Tong}\ \emph {et~al.}(2015)\citenamefont {Tong}, \citenamefont {Tan},\ and\ \citenamefont {Xu}}]{Tong2015}%
  \BibitemOpen
  \bibfield  {author} {\bibinfo {author} {\bibfnamefont {H.}~\bibnamefont {Tong}}, \bibinfo {author} {\bibfnamefont {P.}~\bibnamefont {Tan}}, \ and\ \bibinfo {author} {\bibfnamefont {N.}~\bibnamefont {Xu}},\ }\href {\doibase 10.1038/srep15378} {\bibfield  {journal} {\bibinfo  {journal} {Scientific Reports}\ }\textbf {\bibinfo {volume} {5}},\ \bibinfo {pages} {15378} (\bibinfo {year} {2015})}\BibitemShut {NoStop}%
\bibitem [{\citenamefont {Chhajed}\ \emph {et~al.}(2025)\citenamefont {Chhajed}, \citenamefont {Gruber}, \citenamefont {Dye}, \citenamefont {J\"ulicher},\ and\ \citenamefont {Popovi\'c}}]{Chhajed2025}%
  \BibitemOpen
  \bibfield  {author} {\bibinfo {author} {\bibfnamefont {K.}~\bibnamefont {Chhajed}}, \bibinfo {author} {\bibfnamefont {F.~S.}\ \bibnamefont {Gruber}}, \bibinfo {author} {\bibfnamefont {N.~A.}\ \bibnamefont {Dye}}, \bibinfo {author} {\bibfnamefont {F.}~\bibnamefont {J\"ulicher}}, \ and\ \bibinfo {author} {\bibfnamefont {M.}~\bibnamefont {Popovi\'c}},\ }\href {\doibase 10.48550/arXiv.2505.05437} {\  (\bibinfo {year} {2025}),\ 10.48550/arXiv.2505.05437},\ \bibinfo {note} {arXiv:2505.05437 [physics]}\BibitemShut {NoStop}%
\end{thebibliography}%

\end{document}